\documentclass[aps,prx,reprint,superscriptaddress,amsmath,amssymb]{revtex4-2}
\pdfoutput=1
\usepackage[utf8]{inputenc}
\usepackage[english]{babel}
\usepackage[T1]{fontenc}
\usepackage{amsmath}
\usepackage{amsfonts}
\usepackage{mathtools}
\usepackage{hyperref}
\usepackage{braket}
\usepackage{enumitem}
\usepackage{makecell}
\usepackage{optidef}
\usepackage{nicefrac}
\usepackage{lipsum}
\usepackage{comment}
\usepackage{float}
\usepackage{algorithm}
\usepackage{algpseudocode}
\usepackage{tikz}
\usetikzlibrary{quantikz2}
\usepackage[normalem]{ulem}
\usepackage[bold,full]{complexity}
\usepackage{soul}
\usepackage{subcaption}
\usepackage{float}

\newlang{\MAXQP}{MaxQP}
\newlang{\MAXCUT}{MaxCut}

\tikzset{
  topdashed/.style={append after command={
    [dashed, red, thick] (\tikzlastnode.north west) -- (\tikzlastnode.north east)}},
  botcolor/.style={append after command={
    [blue, thick] (\tikzlastnode.south west) -- (\tikzlastnode.south east)}},
}

\renewcommand{\vec}[1]{\ensuremath{\boldsymbol{#1}}}
\newcommand{\citationneeded}[1][]{\textsuperscript{\color{blue} [citation needed]}}

\definecolor{PythonGreen}{RGB}{44, 160, 44}
\definecolor{PythonOrange}{RGB}{255, 127, 14}
\definecolor{PythonBlue}{RGB}{31, 119, 180}
\definecolor{PythonRed}{RGB}{214, 39, 40}
\definecolor{PythonPurple}{RGB}{148, 103, 189}
\definecolor{PythonBrown}{RGB}{140, 86, 75}

\begin{document}

\title{Problem-adjusted $X$-mixer eigenstates for improved QAOA}
\author{Matvii Kistaiev}
\affiliation{IntellectEU, Belgium}
\affiliation{Department of Mathematical Methods of Information Security, Igor Sikorsky Kyiv Polytechnic Institute, Ukraine}

\author{Georgios Korpas}
\affiliation{Quantum Technologies Group, The Hongkong and Shanghai Banking Corporation, Singapore}


\author{Bart Partoens}
\affiliation{Department of Physics, University of Antwerp, Belgium}

\author{Matthias Van der Donck}
\affiliation{IntellectEU, Belgium}

\begin{abstract}
We show that the use of problem-adjusted excited eigenstates of the $X$-mixer Hamiltonian can be beneficial for the performance of the Quantum Approximate Optimization Algorithm (QAOA) when applied to problems with degenerate ground states. We show that this advantage persists in the presence of quantum gate and readout noise, with the problem-adjusted excited initial states achieving better optimal performance than the standard ground state initialization. The realized advantage depends on the number of QAOA layers and whether or not the initial state has a reasonable time evolution path into the ground space of the problem Hamiltonian. We also explain how to prepare these problem-adjusted $X$-mixer eigenstates exactly, as well as how to approximate them using a linear combination of ladder operators. Finally, we discuss how the power of these initial states can be applied to any problem Hamiltonian regardless of the degeneracy of its ground state by making use of a Grover-style mixer Hamiltonian.
\end{abstract}

\maketitle


\section{Introduction}
\label{introduction}
Despite the early theoretical successes in the field of quantum computing \cite{grover1996fast, shor1999polynomial}, there is still a very active search for additional practically useful quantum algorithms \cite{dalzell2023quantum}. One area in which the applicability of quantum algorithms is still largely unknown is optimization \cite{abbas2023quantum}. For combinatorial optimization problems in particular, considerable attention has been focused on heuristic algorithms, most notably on the Quantum Approximate Optimization Algorithm (QAOA) \cite{farhi2014quantum, hadfield2019quantum, blekos2024review}. This algorithm has been applied to many different problem classes, such as Maximum Cut \cite{farhi2014quantum}, Maximum Independent Set \cite{choi2019tutorial, zhou2020quantum}, Binary Least Squares \cite{borle2021quantum}, and Portfolio Optimization \cite{brandhofer2022benchmarking}, although a fair comparison with classical algorithms is sometimes lacking. Nevertheless, there are indications that QAOA may offer a computational advantage \cite{guo2026setting}, e.g. it has been theoretically shown that the output distribution of QAOA cannot in general be efficiently sampled classically \cite{farhi2016quantum}, and empirical evidence suggests that it can outperform state-of-the-art classical algorithms for sufficiently large problem sizes \cite{shaydulin2023evidence, montanez2025toward}. On the other hand, recent results have also highlighted limitations of shallow and noisy variational quantum circuits. In particular, fixed-depth QAOA training was shown to admit a fully polynomial-time randomized approximation scheme, imposing restrictions on the class of hard optimization problems it can solve \cite{sakos2026globaloptimizationparametrizedquantum}, while uncorrected noise can render typical quantum circuits effectively shallow and enable their efficient classical approximation in certain regimes \cite{mele2026noise}.

By design, QAOA is a variational, hybrid algorithm in which a classical optimizer minimizes a cost function which is evaluated on a quantum computer. This hybrid structure makes it suitable for NISQ devices \cite{preskill2018quantum} but also leads to one of its main limitations, namely the hardness of the classical optimization \cite{bittel2021training} due to, e.g., barren plateaus \cite{mcclean2018barren}. However, this can be somewhat mitigated by making use of iterative optimization schemes and/or lower-dimensional (re-)parametrizations \cite{zhou2020quantum, apte2025iterative}, as well as by reusing parameters across different problem instances \cite{brandao2018fixed, shaydulin2023parameter, sureshbabu2024parameter, montanez2025transfer}. In addition, several other methods for addressing the barren plateaus problem have been proposed in the literature \cite{Mele2022,mastropietro2026parallel,shi2025avoiding}.

Over the past several years, many QAOA variants have been proposed with the aim of improving performance. The vast majority of these modify the parametrized part of the variational ansatz, for example by using different (constraint-preserving) mixer Hamiltonians \cite{hadfield2019quantum, wang2020xy}, leveraging intermediate Hamiltonians \cite{chandarana2022digitized}, warm-starting based on the solution of a continuous relaxation of the problem \cite{egger2021warm}, or increasing the number of variational parameters \cite{herrman2022multi}. Much less attention has been paid to the use of initial states in the QAOA ansatz that are not the ground state of the mixer Hamiltonian. Ref. \cite{crosson2014different} proposed starting in random first excited states of the mixer Hamiltonian in the adiabatic quantum computing framework and reported promising results for hard instances of MAX 2-SAT. However, as we will show, this approach does not make optimal use of the excited-state eigenspace, nor does it provide an intuitive explanation for why this approach can outperform ground state initialization.

Concretely, in this work, we explain when and why it is advantageous to initialize QAOA in an excited state of the $X$-mixer Hamiltonian. We show how to determine the appropriate eigenstates within the degenerate excited eigenspaces using degenerate perturbation theory, and we propose efficient quantum circuits for preparing these states. We present both exact and approximate constructions and as such enable a trade-off between initial-state quality and required quantum hardware resources. We explain how the excited initial states can be used in a Grover-style mixer Hamiltonian to unlock general applicability, and we benchmark our approach on two classes of optimization problems demonstrating that it can significantly outperform ground state initialization both in the absence and presence of quantum gate and readout noise.

The remainder of the paper is organized as follows. In Sec. \ref{backgroundandmethodology}, we detail our methodology. We give an overview of the studied problem classes and their characteristics, describe the parameter optimization scheme used in our QAOA framework, and define the target performance metrics. In Sec. \ref{results}, we present our results. We introduce both our exact and approximate excited-state preparation protocols, and demonstrate their beneficial impact on QAOA performance for problem Hamiltonians with both known and unknown ground state degeneracy. Finally, in Sec. \ref{conclusion}, we summarize our main conclusions.

\section{Background and Methodology}
\label{backgroundandmethodology}

\subsection{Quantum Approximate Optimization Algorithm}
\label{qaoa}
The original \cite{farhi2014quantum}, and still widely used, implementation of QAOA makes use of a variational quantum state
\begin{equation}
\label{eq:qaoa}
\ket{\vec{\gamma}, \vec{\beta}} = \prod_{j=1}^pe^{-i\beta_jH_M}e^{-i\gamma_jH_P}\ket{\psi_0},
\end{equation}
where $\vec{\gamma}=(\gamma_1,\ldots,\gamma_p)$ and $\vec{\beta}=(\beta_1,\ldots,\beta_p)$ are the $2p$ real variational parameters, $H_P$ is the problem Hamiltonian whose ground state is sought, $H_M$ is known as the mixer Hamiltonian, and $\ket{\psi_0}$ is some suitable initial state. The hyperparameter $p$ controls the number of so-called QAOA layers used to prepare $\ket{\vec{\gamma}, \vec{\beta}}$, with higher $p$ values leading to increased state preparation complexity, but also to enhanced expressibility \cite{holmes2022connecting}. In particular, for appropriately chosen Hamiltonians and initial state, it is guaranteed that in the limit $p\rightarrow\infty$ parameters $\vec{\gamma}$ and $\vec{\beta}$ exist such that $\ket{\vec{\gamma}, \vec{\beta}}$ is exactly equal to the ground state of $H_P$ \cite{farhi2014quantum}.

The problem Hamiltonian $H_P$ is often a diagonal Hamiltonian corresponding to a combinatorial optimization problem, although in principle it can also be non-diagonal. We detail our choices of $H_P$ in Sec. \ref{studiedproblems}. The mixer Hamiltonian $H_M$ can in principle also be chosen arbitrarily. However, it is advantageous if $H_M$ has a known ground state that can be efficiently prepared on a quantum computer. In this case, the initial state $\ket{\psi_0}$ in Eq. \eqref{eq:qaoa} can be taken to be this ground state, which typically results in improved and more stable QAOA performance. The most common choice for $H_M$ is the so-called $X$-mixer Hamiltonian $H_M=-\sum_iX_i$. Both the time-evolution operator of this Hamiltonian and the preparation of its ground state $\ket{+}^{\otimes n}$, where $n$ is the number of qubits, can be implemented using constant-depth quantum circuits consisting exclusively of single-qubit quantum gates. Here, we also mainly focus on this mixer Hamiltonian; however, we will show that initializing QAOA in certain excited states of $H_M$ instead of the ground state can be advantageous. In addition, we will show how a Grover-style mixer Hamiltonian based on these excited states can unlock further performance improvements.

\subsection{Alternative initial states}
\label{alternativeinitialstates}
QAOA can be related to the time evolution generated by the Hamiltonian
\begin{equation}
    \label{eq:timehamiltonian}
    H(t) = \beta(t)H_M+\gamma(t)H_P,
\end{equation}
where $\beta(t)$ and $\gamma(t)$ are functions that smoothly transition from 1 to 0 and from 0 to 1, respectively, as $t$ increases. In particular, a suitable choice of QAOA angles yields a Trotterized and time-discretized approximation of this evolution. If the system is initialized in the ground state of $H_M$ and the total evolution time is sufficiently long, then the continuous-time evolution becomes adiabatic and the system remains in the instantaneous ground state, ultimately reaching the ground state of $H_P$ \cite{farhi2001quantum, albash2018adiabatic}. For QAOA to accurately approximate this adiabatic evolution, the number of layers must be sufficiently large to provide a fine-grained time discretization. Alternatively, QAOA can also be interpreted as an application of `bang-bang' quantum control, which is optimal for a fixed computation time \cite{yang2017optimizing}. However, several examples have been discussed where adiabatic quantum computing \cite{somma2012quantum} or QAOA \cite{zhou2020quantum, wang2025exponential} are able to bypass the adiabatic restrictions resulting from the exponentially small gap between the lowest two energy states by temporarily transitioning out of the ground state before returning back to it. QAOA, in particular, is well-suited for leveraging these kinds of non-adiabatic effects, because it is able to find the (locally) optimal, possibly non-linear, interpolation schedule, even when the total evolution time is limited. This schedule flexibility also enables QAOA to potentially achieve good overlap with the ground state of $H_P$ even when starting in alternative, e.g. excited, initial states. This is particularly relevant if the ground state of $H_P$ is $d$-fold degenerate, since in this case there are $d$ eigenstates of $H_M$ which can be adiabatically evolved into the desired ground space. Which eigenstates of $H_M$ end up in the ground space depends on the properties of $H_P$ and the conserved symmetries along the evolution path.

\subsection{Studied problems}
\label{studiedproblems}
In order to explore this idea of exploiting the degeneracy of the ground space of the problem Hamiltonian, we consider two Ising problem classes with different ground state degeneracies. The first is the Sherrington--Kirkpatrick (SK) model, which we define as
\begin{equation}
\label{eq:skhamiltonian}
\begin{split}
&H_P^{\mathrm{SK}}=\frac{1}{\sqrt{n}}\sum_{i<j}^nJ_{ij}Z_iZ_j, \\
&J_{ij}\sim\mathcal{N}(0,1).
\end{split}
\end{equation}
The SK Hamiltonian is invariant under global spin flip $\prod_iX_i$ \cite{larocca2022diagnosing}, which maps each computational basis state to its spin-flipped counterpart. Since the couplings are drawn from a continuous distribution, accidental degeneracies between spin configurations not related by symmetry occur with probability zero and, consequently, a typical SK instance has a two-fold degenerate ground space generated only by this spin-flip symmetry.

The second problem class is the Circulant Ising (CI) model, defined as
\begin{equation}\label{eq:cihamiltonian}
\begin{split}
&H_P^{\mathrm{CI}}=\frac{1}{\sqrt{n}}\sum_{i<j}J_{\delta(i,j)} Z_iZ_j\,,\\
&\delta(i,j) = \min\big(\vert i-j\rvert,\, n - \lvert i-j\rvert\big),\\
&J_{\delta(i,j)}\sim\mathcal{N}(0, 1).
\end{split}
\end{equation}
Here, the coupling depends only on the distance $\delta$ between spins on a ring. In addition to the global spin flip, the Hamiltonian is invariant under translations $T\ket{s_1s_2\ldots s_n}=\ket{s_ns_1\ldots s_{n-1}}$ and site reversal $R\ket{s_1s_2\ldots s_n}=\ket{s_ns_{n-1}\ldots s_1}$. These operations generate the dihedral symmetries of the ring, and together with the global spin flip they can produce a ground space containing up to $4n$ computational basis states. This degeneracy is reduced whenever a ground state configuration has a nontrivial stabilizer under these symmetries. For example, if the reflected configuration is related to the original one by a translation, then reflection does not enlarge the translational orbit, resulting in a $2n$-fold degeneracy when the spin-flipped orbit remains distinct. Further reductions can occur in the presence of additional stabilizers, particularly for non-prime $n$, where configurations can exhibit a shorter spatial period. However, in this work we will only consider the $4n$-degenerate cases to demonstrate our approach as this is the dominant subclass of problems in the large $n$ limit. Note that the mixer Hamiltonian $H_M$ commutes with all three of these symmetries, so they correspond to conserved quantities along the adiabatic evolution trajectory.

\subsection{Parameter optimization scheme}
\label{parameteroptimizationscheme}
We obtain the variational parameters $\vec{\gamma}$ and $\vec{\beta}$ by numerically minimizing the expectation value of $H_P$ with respect to the variational state \eqref{eq:qaoa}, i.e., $\bra{\vec{\gamma}, \vec{\beta}}H_P\ket{\vec{\gamma}, \vec{\beta}}$, on a classical computer. However, as $p$ grows, optimizing all $2p$ parameters at once over an unknown and nontrivial landscape becomes increasingly difficult. Several more efficient strategies have therefore been proposed, such as the \textsc{Interp} and \textsc{Fourier} schemes of Zhou et al.~\cite{zhou2020quantum} and its generalization, the Iterative Interpolation (II) scheme of Apte et al.~\cite{apte2025iterative}. These methods rely on the observation that the optimal parameters $\vec\gamma$ and $\vec\beta$ are usually not independent, but rather follow some smooth curves $\gamma(s)$ and $\beta(s)$ along the adiabatic path $0\leq s\leq1$. The $p$ angles of $\vec\gamma$ (and likewise $\vec\beta$) are then obtained by sampling the corresponding curve at $p$ evenly-spaced points, and each curve can typically be described by only $q\leq p$ parameters. Optimization is therefore carried out over these $q$ curve parameters rather than the $p$ angle parameters, which reduces the search space's dimensionality. Another closely related observation is that the underlying evolution curves change little when going from $p$ to $p+1$ layers (for sufficiently large $p$). These schemes exploit this fact to grow the QAOA's circuit one layer at a time, using the curve fitted at depth $p$ as initial guess for depth $p+1$, significantly reducing the total optimization hardness. In this work, we use the II scheme of Ref. \cite{apte2025iterative} with Chebyshev polynomials as an orthonormal functional basis for the curve parametrization.

The quality of our obtained solutions is measured by means of two different metrics: the ground space probability
\begin{equation}
\label{gs_overlap}
P_{GS} = \lVert \Pi_{GS}\ket{\vec{\gamma}^*,\vec{\beta}^*}\rVert^2,
\end{equation}
where $\vec{\gamma}^*$ and $\vec{\beta}^*$ are the optimized variational parameters and $\Pi_{GS}$ is the projector onto the (possibly degenerate) ground state eigenspace of $H_P$, and the approximation ratio
\begin{equation}
\label{approxratio}
AR = \frac{\bra{\vec{\gamma}^*,\vec{\beta}^*} H_P\ket{\vec{\gamma}^*,\vec{\beta}^*}}{C_{min}},
\end{equation}
where $C_{min}$ is the smallest eigenvalue of $H_P$. With this definition, an $AR$ value of $1$ corresponds to solving the problem exactly, and for the problem types considered in this work an $AR$ value of $0$ corresponds to random guessing.

\section{Results}
\label{results}
As mentioned in Sec. \ref{alternativeinitialstates}, it can be of interest to initiate QAOA in an excited state of the initial Hamiltonian. We will now discuss how to find and construct advantageous initial states of the $X$-mixer Hamiltonian and evaluate their performance in QAOA.
\begin{figure*}[t]
\centering
\begin{subfigure}[c]{0.475\textwidth}
    \includegraphics[width=7.75cm]{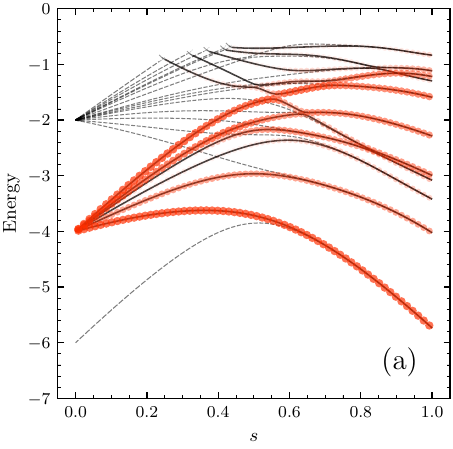}
\end{subfigure}
\hfill
\begin{subfigure}[c]{0.475\textwidth}
\includegraphics[width=7.75cm]{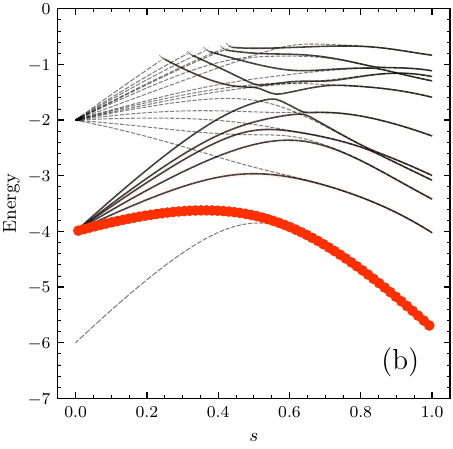}
\end{subfigure}
\caption{The lowest few energy levels for a linear interpolation between the $X$-mixer Hamiltonian ($s=0$) and an instance of the SK Hamiltonian ($s=1$) for $n=6$. Solid (dashed) lines indicate states having an eigenvalue of $-1$ ($+1$) with respect to the global spin flip operator, i.e. they indicate 2 symmetry sectors which cannot mix during evolution. The opacity and thickness of the red circles indicate the fraction of the state population in each of the instantaneous energy levels while the state is being evolved near-adiabatically after being initialized in (a) a naive product state $\ket{+-++++}$ within the first excited eigenspace and (b) the 0-th problem-adjusted first excited energy level eigenstate of the $X$-mixer Hamiltonian $\ket{\alpha_1^0}$.}
\label{fig:energy_bands}
\end{figure*}

\subsection{Problem-adjusted excited eigenstates of the \mbox{$X$-mixer} Hamiltonian}
\label{exactexcitedstatepreparationprotocol}
\begin{figure*}[t]

\tikzset{
  uc/.pic={
    \begin{pgfinterruptboundingbox}
      \draw[line width=0.5pt]
        (-0.35em,-0.35em) -- (0,0) -- (0.35em,-0.35em);  
      \fill (0,0) circle[radius=0.12em];                   
    \end{pgfinterruptboundingbox}
  }
}

\tikzset{
  dc/.pic={
    \begin{pgfinterruptboundingbox}
      \draw[line width=0.5pt]
        (-0.35em,0.35em) -- (0,0) -- (0.35em,0.35em);  
      \fill (0,0) circle[radius=0.12em];                   
    \end{pgfinterruptboundingbox}
  }
}

\NewExpandableDocumentCommand{\onwire}{O{} m}{%
  |[label={#1}, append after command={(\tikzlastnode.center) pic{#2}}]| {}}
  
\begin{center}
  \begin{subfigure}[b]{0.525\textwidth}
    \centering
    \begin{quantikz}[column sep={24pt,between origins}, row sep={18pt,between origins}]
        \lstick{$\ket{0}$}&\gate{X}&\onwire[{[label distance=0.4em] above:\theta_1}]{uc}\vqw{5}& \control{}\vqw{5} & \control{}\vqw{5} & \control{}\vqw{5} &\onwire[{[label distance=0.4em] above:\theta_5}]{uc}\vqw{5} &                   && \rstick[7]{$\ket{\psi}$} \\
        \lstick{$\ket{0}$}&\gate{X}& \control{}       & \control{}        & \onwire[{[label distance=0.4em] above right:\theta_3}]{uc}       &                   &                   &                   &&  \\
        \lstick{$\ket{0}$}&        &                  &                   & \onwire{dc}       & \onwire[{[label distance=0.4em] above right:\theta_4}]{uc}       & \control{}        & \control{}\vqw{2} &\control{}\vqw{4}&  \\
        \lstick{$\ket{0}$}&        &                  & \onwire[{[label distance=0.4em] above right:\theta_2}]{uc}       &                   & \onwire{dc}       & \onwire{dc}       & \control{}        &\control{}&  \\
        \lstick{$\ket{0}$}&        & \onwire{dc}      &                   &                   &                   &                   & \swap{1}          &\onwire[{[label distance=0.4em] above right:\theta_6}]{uc}&  \\
        \lstick{$\ket{0}$}&\gate{X}& \control{}       & \onwire{dc}       & \control{}        & \control{}        & \control{}        & \swap{0}          &&  \\
        \lstick{$\ket{0}$} & & & & & & & & \onwire{dc} &
        \end{quantikz}

    \subcaption{}
    \label{fig:exact_circ}
  \end{subfigure}
  \hfill
  \begin{subfigure}[b]{0.425\textwidth}
    \centering
    \begin{tikzpicture}
        \node[circle, draw=black, fill=black, inner sep=2pt, label=$0100110$] (n1) at (0, 2.5) {};
        \node[circle, draw=black, fill=black, inner sep=2pt, label=$1100010$] (n2) at (1.65, 2.15) {};
        \node[circle, draw=black, fill=black, inner sep=2pt, label=below:$1101000$] (n3) at (0.3, 0.85) {};
        \node[circle, draw=black, fill=black, inner sep=2pt, label=$1010010$] (n4) at (3.5, 2.25) {};
        \node[circle, draw=black, fill=black, inner sep=2pt, label=$1001010$] (n5) at (5.5, 2.75) {};
        \node[circle, draw=black, fill=black, inner sep=2pt, label=below:$0011100$] (n7) at (5, 0.75) {};
        \node[circle, draw=black, fill=black, inner sep=2pt, label=below left:$0011001$] (n8) at (3, 0.8) {};
        
        \draw[line width=1pt] (n1) -- (n2) node[midway, below] {\footnotesize $1$};
        \draw[line width=1pt] (n2) -- (n3) node[pos=0.4, below=0.4em] {\footnotesize $1$};
        \draw[line width=1pt] (n2) -- (n4) node[pos=0.625, below] {\footnotesize $1$};
        \draw[line width=1pt] (n4) -- (n5) node[midway, above] {\footnotesize $1$};
        \draw[line width=1pt] (n4) -- (n7) node[pos=0.25, below=0.5em] {\footnotesize $2$};
        \draw[line width=1pt] (n7) -- (n8) node[pos=0.65, below=0.2em] {\footnotesize $1$};
    
        \node[circle, draw=blue, line width=1.25pt, inner sep=1.25pt,           label=above right:$0011010$] (n6) at (4.3, 1.45) {};
    \end{tikzpicture}
    \vspace{10px}
    \caption{ }
    \label{fig:mst}
  \end{subfigure}
  \end{center}
    \caption{(a) An example-circuit that prepares a $7$-qubit state $\ket{\psi}$ with fixed Hamming weight of $3$. (b) A Minimum Spanning Tree of a graph of the basis vectors that have non-zero amplitudes in $\ket{\psi}$ with weights being the SWAP-distance between the bitstrings. For this example we consider the state $\ket{\psi}$ to be
    \makebox[\linewidth]{\raisebox{0pt}[13pt][10pt]{$\displaystyle \ket{\psi} = \alpha_1\ket{{0100110}} + \alpha_2\ket{1100010} + \alpha_3\ket{1010010} + \alpha_4\ket{1101000} + \alpha_5\ket{1001010} + \alpha_6\ket{0011100} + \alpha_7\ket{0011001}$}},
    where all amplitudes are considered to be real-valued.
    We construct $\ket{\psi}$ starting from the basis state $\ket{1100010}$. The rotation angles $\theta_1,\ldots,\theta_6$ are calculated as $\pm\arcsin(\alpha_t/\alpha_s)$, where $\alpha_s$ is the current amplitude of the source basis state, $\alpha_t$ is the $L^2$-norm of the target amplitudes in the remaining branch of the tree, and the sign depends on whether the source state is $\ket{01}$ or $\ket{10}$. The controlled SWAP gate is a simplified Givens rotation with $\theta = \pi/2$; it appears because we pass through the intermediate state $\ket{0011010}$ (blue dot), which has zero amplitude in $\ket{\psi}$.}
  \label{fig:exact_state_prep}
\end{figure*}
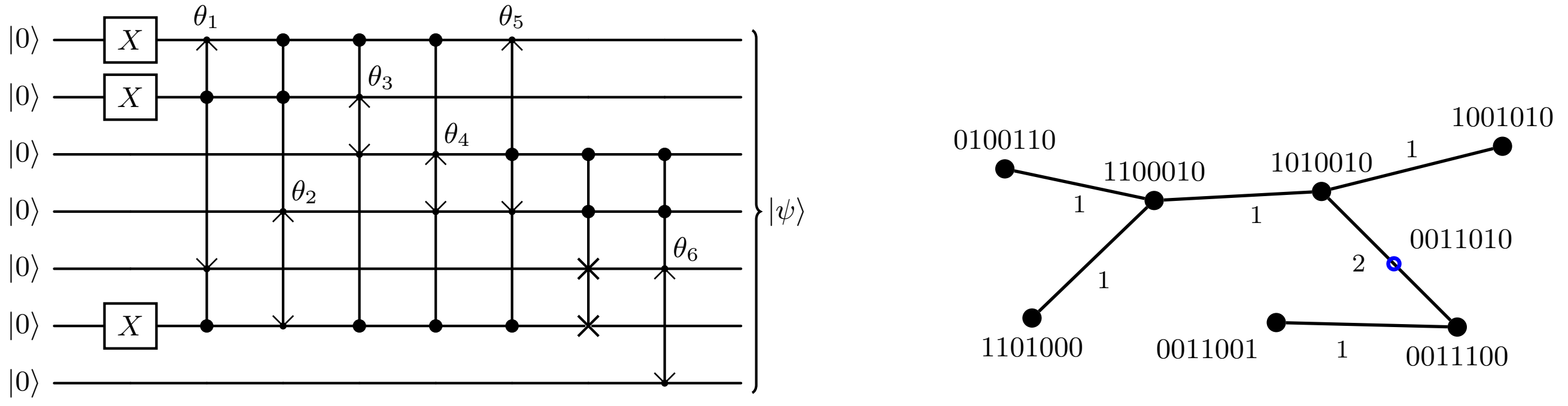

The energy spectrum of the $n$-qubit $X$-mixer Hamiltonian consists of $n+1$ distinct energy levels, where the $k$-th distinct energy level (with $k=0$ corresponding to the ground state) corresponds to a ${n\choose k}$-dimensional degenerate subspace that we denote  $X^n_k$. The most straightforward choice for an orthonormal basis for this subspace is
\begin{equation}
\label{subspacebasis}
B^n_k =  \big\{ H^{\otimes n}\ket{x}\, : \; x\in\{0, 1\}^n  \,\text{ and }\, wt(x) = k\big\}\,,
\end{equation}
where $wt(x)$ is the Hamming weight of $x$. Such excited states are easy to prepare on a quantum computer, however in general adiabatically evolving them along a given schedule will not track the instantaneous eigenstates of the total Hamiltonian~\eqref{eq:timehamiltonian}, but rather some linear combination thereof determined by the specifics of $H_P$. This is illustrated in Fig. \ref{fig:energy_bands}(a), which shows how a naive product state within the first excited subspace near-adiabatically evolves into a multitude of problem Hamiltonian eigenstates, including some highly excited ones.

The correct way to determine the excited initial states which will track the instantaneous eigenstates for infinitesimally small $s>0$ is by making use of degenerate perturbation theory. More specifically, the $l$-th instantaneous eigenstate originating from the $k$-th excited eigenspace of $H_M$ corresponds to the $l$-th-eigenvalue eigenstate of the perturbation operator $(\beta(t_1)-1)H_M+\gamma(t_1) H_P$, for some small $t_1$, within the $k$-th excited eigenspace. We call such a state the $l$-th problem-adjusted $k$-th energy level eigenstate of the $X$-mixer Hamiltonian and denote it $\ket{\alpha_k^l}$. Note that if $H_P$ is real-valued, as is the case for the diagonal Hamiltonians considered in this work, then the coefficients of $\ket{\alpha_k^l}$ expressed in the basis $B_k^n$ are also real-valued. Fig. \ref{fig:energy_bands}(b) shows that such a state indeed tracks an individual energy level under (near-)adiabatic evolution. If only a subset of the energy levels of a certain initial subspace end up in the final ground space, then the initial state $\ket{\alpha_k^0}$ typically has a clear advantage compared to the naive initial states, such as the one shown in Fig. \ref{fig:energy_bands}(a). Even if all levels of a subspace end up in the final ground space, an initial state $\ket{\alpha_k^l}$ with $l$ small compared to ${n\choose k}$ may still be beneficial since it is less likely to transition to higher energy levels under non-adiabatic evolution. On average, initial states with lower values of $l$ are expected to evolve into final states with lower energy, however this may not be true for each individual problem instance, e.g. due to (avoided) crossings in the evolution path. In all of our experiments whose results are presented in this work, we have set $l=0$.

In the $X^n_k$ eigenspace the $H_M$ operator acts as the identity operator, therefore diagonalization of the perturbation operator reduces to just diagonalizing $H_P$ restricted to the $X^n_k$ subspace. In general, for a Hamiltonian with a polynomial number of terms $\#H_P$, the restricted operator's $H_P \vert_{X^n_k}$ matrix in the basis $B_k^n$ can be constructed classically with complexity $\mathcal{O}\left(\#H_P\cdot n^{k+2}\right)$ and diagonalized in $\mathcal{O}(n^{3k})$. This quickly becomes practically forbidding as $k$ increases. However, since the basis size increases exponentially with $k$ while each of the terms of $H_P$ maps each basis state to at most 1 other basis state, the matrix becomes increasingly sparse for higher excited subspaces with at most $\binom{n}{k} \cdot \min\{\binom{n}{k},\, \#H_P\} $ non-zero entries. In particular, for the $2$-local SK and CI Hamiltonians that we consider, the matrix of $H_P \vert_{X^n_k}$ has sparsity at least $1 - {(n-k)k} / {n \choose k}$, allowing for effectively much faster computations. Furthermore, we are only interested in the $l+1$ lowest-energy eigenvectors of $H_P \vert_{X^n_k}$, which admits an additional computational speedup using for example the Lanczos method. The combination of the lowest-energy sector and sparsity arguments makes the computation of $\ket{\alpha_k^l}$ practically tractable for considerably larger system sizes than the worst-case scaling might suggest. For example, for the CI problem with $n=100$, $k=3$, and $l=0$, the problem-adjusted excited eigenstate can be computed on a laptop equipped with an Apple M1 processor and 16 GB of memory in 3-4 minutes.

If after the diagonalization the $l$-th eigenstate of $H_P \vert_{X^n_k}$ is still part of a degenerate subspace within first order perturbation theory, then we would need to resort to higher order perturbations. These require summations over matrix elements of $H_P$ involving all of the exponentially many eigenstates, however only a polynomial number of them are non-zero and therefore these higher order contributions can still be efficiently computed classically. For some instances of certain problem classes, some eigenstates of $H_P \vert_{X^n_k}$ can remain degenerate along the entire evolution path from $H_M$ to $H_P$. In such a case, it is not possible to separate the states using perturbation theory, but it is also not needed. Finally, note that even though in this work we only consider problem classes for which $H_P$ is diagonal, the procedure described above works for non-diagonal problem Hamiltonians as well.

\subsubsection{Exact state preparation}
\label{exactstatepreparation}
\begin{figure*}[t]
\centering
\begin{subfigure}[c]{0.475\textwidth}
    \includegraphics[width=7.75cm]{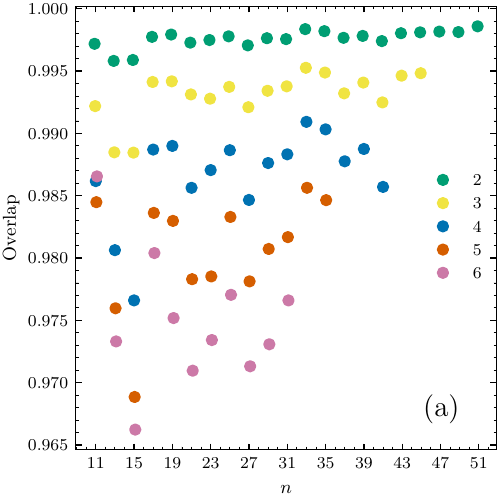}
\end{subfigure}
\hfill
\begin{subfigure}[c]{0.475\textwidth}
\includegraphics[width=7.75cm]{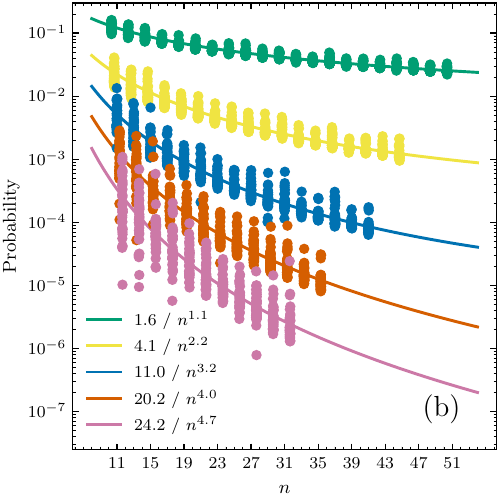}
\end{subfigure}

\caption{(a) Median overlap between the approximate excited state $\ket{\widetilde{\alpha}_k^0}$ and the exact excited state $\ket{\alpha_k^0}$ for $k$ ranging from $2$ to $6$ as a function of $n$ for $50$ instances per problem size of the CI Hamiltonian. (b) Corresponding success probabilities of preparing the approximate state $\ket{\widetilde{\alpha}_k^0}$ along with their hyperbolic function fits for each $k$.}
\label{fig:init_state_approx}
\end{figure*}

In order to use the described problem-adjusted excited eigenstates of the $X$-mixer Hamiltonian in QAOA, we need an efficient quantum circuit that prepares $\ket{\alpha_k^l}$. The following construction for fixed Hamming weight state preparation combined with a Hadamard layer $H^{\otimes n}$ at the end can be used to prepare $\ket{\alpha_k^l}$ using $\mathcal{O}(n^k)$ single-qubit and CNOT gates.

\textit{Construction 1.} Given an $n$-qubit state $\ket{\psi}$ that is a superposition of $m$ basis states of fixed Hamming weight $k$, there exists an $n$-qubit quantum circuit that prepares $\ket{\psi}$ using at most $8\min{\big(m\cdot\min(n,\, n-k),\: {n\choose k}\big)}$ single-qubit unitary and $k$-controlled $X$ gates.

This construction closely follows the idea of Ref. \cite{farias2025hammingencoder}. It relies on the fact that if $\ket{x_i}$ and $\ket{x_j}$ are two computational basis states that differ in exactly two qubits, and $c_i$ and $c_j$ are their amplitudes in the decomposition $\ket{\psi} = \sum_{i=1}^m c_i \ket{x_i}$, then a complex \textsc{RBS} gate, as defined in Eq. (2) of Ref. \cite{farias2025hammingencoder}, applied to the two differing qubits and controlled on the qubits which are equal to $1$ in both $\ket{x_i}$ and $\ket{x_j}$, transfers amplitude between $c_i$ and $c_j$ by any desired amount. Note that for real amplitudes, as is the case in this work, the complex \textsc{RBS} gate simply reduces to a Givens rotation in the $\{\ket{01},\ket{10}\}$ subspace. The state $\ket{\psi}$ is prepared by starting from a single computational basis state with the desired Hamming weight and repeatedly redistributing amplitude to neighboring basis states, choosing each rotation angle so that every state ends up with its target amplitude. For a target state with an arbitrary set of $m \leq \binom{n}{k}$ nonzero amplitudes, the optimal ordering of basis states to which amplitudes are being redistributed is no longer the Gray code ordering used in Sec. III.B of Ref. \cite{farias2025hammingencoder}. Instead, the optimal ordering is obtained by traversing, from root to leaves, a minimum Steiner tree in the Johnson's graph $J(n,k)$ spanning the $m$ bitstrings that appear in the decomposition of $\ket{\psi}$. Note that even though computing an exact minimum Steiner tree is \textsc{NP}-complete, it can be approximated to within a constant factor of $2$ by taking a Minimum Spanning Tree (MST) of the complete weighted graph whose vertices are all $m$ target bitstrings, with each edge weighted by the number of swaps needed to turn one bitstring into the other, which can be computed in $\mathcal{O}(m^2\log m)$. If an edge of weight $t \geq 2$ is a part of the MST, then this amplitude transfer is performed by $t-1$ moves through the intermediate basis states that are not a part of the target state $\ket{\psi}$. An example of such a construction is presented in Fig. \ref{fig:exact_state_prep}.

\subsubsection{Approximate state preparation}
\label{approximatestatepreparation}
\begin{figure*}[t]
\centering

\begin{quantikz}[column sep=7pt, row sep={14pt,between origins}]
    \lstick[3]{\rotatebox{90}{address}} &[14pt] \lstick{$\ket{0}$}\wireoverride{n} && \gate[3]{V_{\vec{d}}} &&     & &&&&&&&\ctrl{5}& &     && &&&     & &\ctrl{5}&&&&&&& &     && \gate[3]{V_{|\vec{d}|}^\dagger} && \meter[4]{\ket{0000}} \wire[d][12]{c}\\
    &[14pt] \lstick{$\ket{0}$}\wireoverride{n} && &&     & &&&&&\ctrl{6}&\ctrl{5}&& &     && &&&     & &&\ctrl{5}&\ctrl{6}&&&&& &     && && \\
    &[14pt] \lstick{$\ket{0}$}\wireoverride{n} && &&     & &\ctrl{9}&\ctrl{8}&\ctrl{7}&\ctrl{6}&&&& &     && &&&     & &&&&\ctrl{6}&\ctrl{7}&\ctrl{8}&\ctrl{9}& &     && && \\[{16pt,between origins}]
    \lstick{sum} &[14pt] \lstick{$\ket{0}$}\wireoverride{n} &&\gate{H}& &    & &&&&&&&& &     && \ctrl{1} &&&      & &&&&&&&& &     & &\gate{H} && \\[{16pt,between origins}]
    \lstick[8]{$\ket{\psi}$} &&& &&     & &&&&\swap{4}& &\swap{2}& \swap{1}&&     && \targ{} & \gate{Z}&&     & &\swap{1} &\swap{2}&  &\swap{4}&&&&&    && & && \rstick[8]{\rotatebox{90}{$\big(\sum_id_{i}L_i^{\dag}\big)\ket{\psi}$}}\\
    &&& &&     & &&&\swap{4}&& \swap{2}&& \swap{-1}&&     &&& &&     & &\swap{-1} &&\swap{2} &&\swap{4}&&&&     && & &&\\
    &&& &&     & &&\swap{4}&&& &\swap{-2}& &&     &&& & &    & & &\swap{-2}& &&&\swap{4}&& &    && & &&\\
    &&& &&     & &\swap{4}&&&& \swap{-2}&& &&     &&& &&     & & &&\swap{-2} &&&&\swap{4}&&     && & &&\\
    &&& &&     & &&&&\swap{-4}& && &&     &&& &&     & & && &\swap{-4}&&&&&     && & &&\\
    &&& &&     & &&&\swap{-4}&& && &&     &&& &&     & & && &&\swap{-4}&&&&     && & &&\\
    &&& &&     & &&\swap{-4}&&& && &&     &&& &&     & & && &&&\swap{-4}&&&     && & &&\\
    &&& &&     & &\swap{-4}&&&& && &&     &&& &&     & & && &&&&\swap{-4}&&     && & &&\\[{16pt,between origins}]
    \setwiretype{c} &&& &    & &&&&&&&& &     && &&&    & &&&&&&&& &    && && &&\rstick{$0000$}
\end{quantikz}

\caption{Circuit implementing a linear combination of raising operators using $\lceil\log_2n\rceil+1$ ancilla qubits for $n=8$.}
\label{fig:laddercircconstr}
\end{figure*}
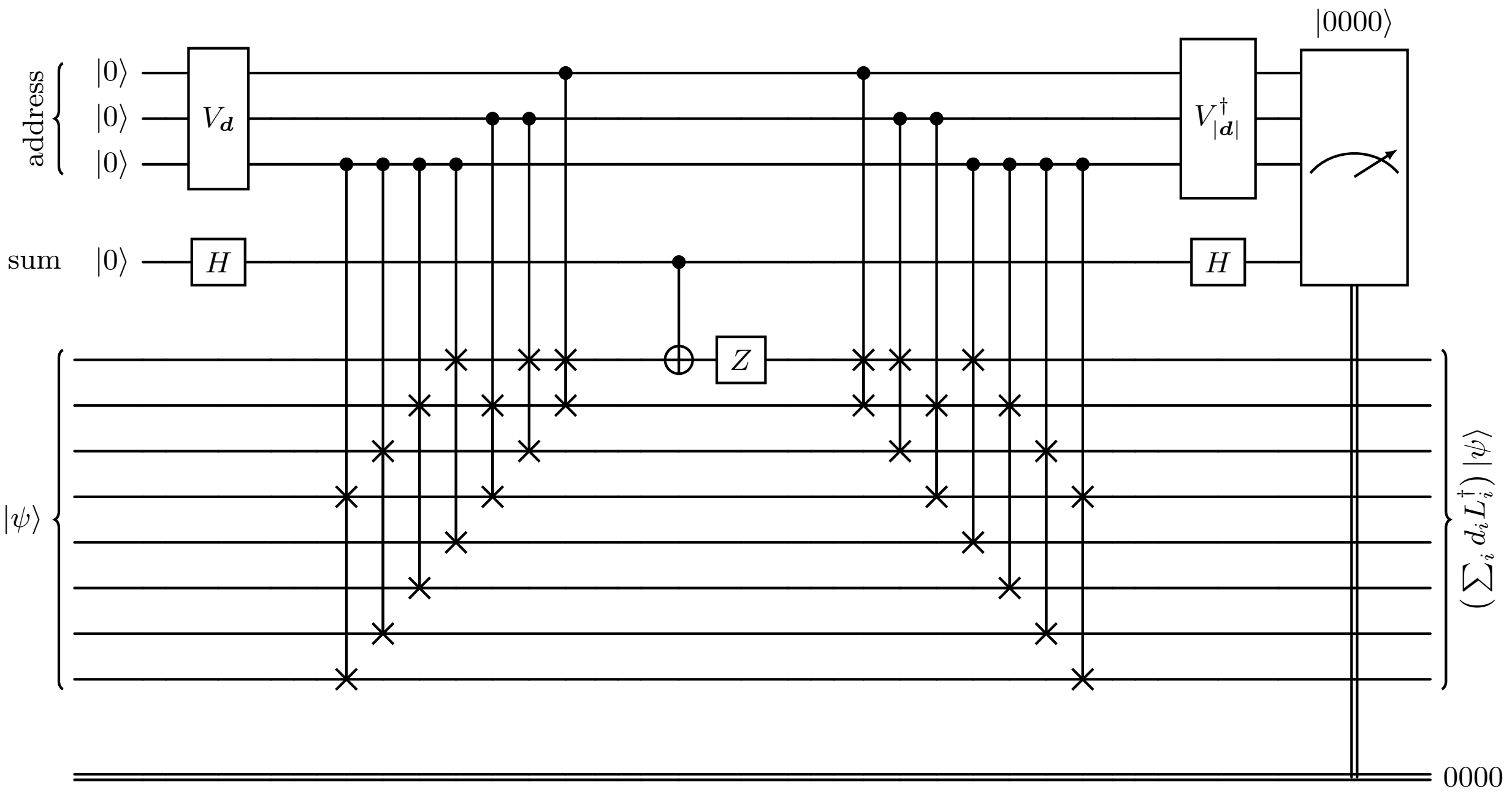

The procedure described in Sec. \ref{exactstatepreparation} for constructing the problem-adjusted excited states $\ket{\alpha_k^l}$, while still polynomial in $n$, quickly becomes prohibitively expensive for large $n$ when $k>1$. However, it is possible to construct an approximation of $\ket{\alpha_k^l}$ using a quantum circuit whose gate count is still linear in $n$. The core idea of this construction is based on the fact that the operator $L^{\dag} = Z+iY=Z+ZX$ is a raising operator for the single qubit $X$-mixer. That is, it obeys the commutation relation $[-X, L^{\dag}]=2L^{\dag}$ and it acts on the eigenstates of the $X$-mixer as $L^{\dag}\ket{+}=2\ket{-}$ and $L^{\dag}\ket{-}=0$. Naturally, its Hermitian conjugate $L=Z-iY$ acts as the corresponding lowering operator. Therefore, for the $n$-qubit $X$-mixer Hamiltonian $H_M$, any operator of the form $\sum_id_iL^{\dag}_i$, with $L^{\dag}_i$ the raising operator acting on qubit $i$ and with $d_i$ free parameters, is a valid raising operator.

Our scheme for approximately constructing the $l$-th level of the $k$-th excited band makes use of the state
\begin{equation}
\label{eq:ladderstate}
\ket{\widetilde{\alpha}_k^l}=\mathcal{C}\prod_{j=1}^k\left(\sum_id_{i,j}L_i^{\dag}\right)\ket{\alpha_0},
\end{equation}
where $\ket{\alpha_0}\equiv\ket{+}^{\otimes n}$, $\mathcal{C}$ is a suitable normalisation constant, and the coefficients $d_{i,j}$ are determined by classically optimizing some cost function. Concretely, we choose to maximize the overlap with the corresponding exact state:
\begin{equation}
    \label{eq:laddercost}
    \begin{split}
    &\underset{\vec{d}_{1},\ldots,\vec{d}_{k}}{\text{max}}\ f_k^l(\vec{d}_{1},\ldots,\vec{d}_{k}) = \lvert\braket{\widetilde{\alpha}_k^l | \alpha_k^l}\rvert 
    \end{split},
\end{equation}
where the vectors $\vec{d}_{j}$ contain the coefficients $d_{0,j},\ldots,d_{n-1,j}$. For a precomputed exact state $\ket{\alpha_k^l}$, e.g. obtained by means of the procedure explained above, this cost function can be classically evaluated in $\mathcal{O}(n^k)$. Alternatively, for $l=0$ one could minimize the expectation value of the problem Hamiltonian $H_P$ with respect to the ladder state \eqref{eq:ladderstate}, which does not require to compute the exact state $\ket{\alpha_k^l}$, however the complexity of evaluating such a cost function is $\mathcal{O}\left(\#H_P\cdot n^{k}\right)$ with $\#H_P$ the number of terms in $H_P$. In both cases the cost function is multilinear, therefore fixing the $k-1$ coefficient vectors $\vec{d}_{j}$ for all $j\neq t$ turns the cost function into a generalized Rayleigh quotient, such that the optimal coefficients $\vec{d}_t$ follow analytically from a generalized eigenvalue problem of size $n$. Sweeping through $t=1,\ldots,\,k$ and repeating until convergence then results in a Block Coordinate Ascent optimization approach that requires $\mathcal{O}(kn)$ cost function evaluations per sweep and, in our experiments, converges quickly and reliably.

The results in Fig. \ref{fig:init_state_approx}(a) show that the approximate excited state preparation protocol achieves quantum states which are very close to the exact excited states, i.e. for the reported values of $n$ and $k$ the median overlaps are all (much) larger than $0.965$. The overlap values decrease slowly with increasing $k$ and even seem to be very slowly increasing with increasing $n$. This is quite remarkable since we are approximating a quantum state which in general consists of ${n \choose k}$ independent coefficients by a quantum state which is defined by only $kn$ parameters. This is also why throughout this work we omit ladder-related results for $k=1$ because in this case our ladder operator ansatz \eqref{eq:ladderstate} becomes equivalent to the exact first excited state $\ket{\alpha^l_1}$. Note that our choice of ansatz could be generalized to any other linear combination of products of $k$ distinct single-qubit ladder operators, possibly admitting more variational parameters and as such potentially achieving even better approximations to the exact quantum states. Here, we focus on our factorized ansatz, for which we now introduce an efficient quantum circuit implementation.
\begin{figure*}[t]
\centering
\begin{subfigure}[c]{0.475\textwidth}
\includegraphics[width=7.75cm]{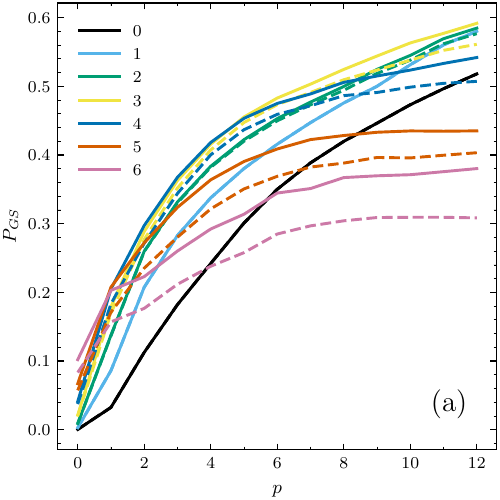}
\end{subfigure}
\hfill
\begin{subfigure}[c]{0.475\textwidth}
\includegraphics[width=7.75cm]{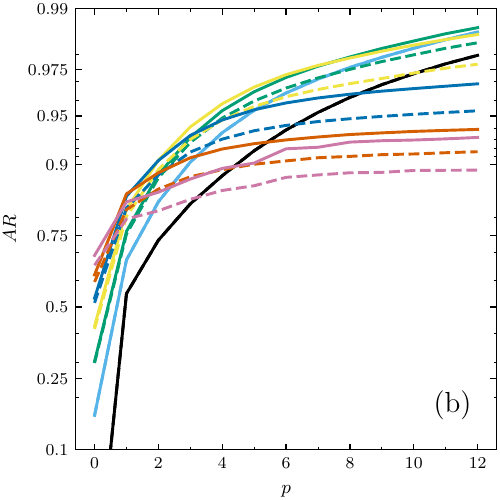}
\end{subfigure}
\caption{(a) Mean $P_{GS}$ of QAOA with $p$ layers for exact initial states $\ket{\alpha_k^0}$ (solid lines) and ladder-approximated initial states $\ket{\widetilde{\alpha}_k^0}$ (dashed lines) for $k$ ranging from $0$ to $6$, averaged across 50 instances of the CI Hamiltonian with $4n$-degenerate ground space for $n=17$. (b) Corresponding mean $AR$ for the same experiments (with $Y$ axis in logit scale for improved readability).}
\label{fig:circ_Ising_perf}
\end{figure*}

\textit{Construction 2.} An $n$-qubit state~$\ket{\widetilde{\alpha}_k^l}$, as defined in Eq. \eqref{eq:ladderstate}, can be prepared probabilistically in an $n$-qubit quantum register using $\lceil\log_2n\rceil + 1$ ancilla qubits and $\mathcal{O}(kn)$ single-qubit and CNOT gates.

This construction can be implemented by means of the LCU protocol defined in \cite{childs2017quantum} since each operator $\sum_id_{i,j}L_i^{\dag}$ for $j=1,\,\ldots\ k$ in its definition is a linear combination of $2n$ unitaries. In general, this protocol implements an operator $\sum_ic_iU_i$ by encoding the square roots of the coefficients (which are assumed to be positive because the phase can be absorbed into the corresponding unitary) in an $n_{\text{addr}}=\lceil\log_2n\rceil$ qubit address register, then applying the operator $\sum_i\ket{i}\!\bra{i} \otimes U_i$ which applies each unitary conditioned on a specific value in the address register, and finally uncomputing the address register. Upon measurement of the all-zeroes bitstring in the address register, the target register then contains the desired state. A concrete circuit implementation for our linear combination of ladder operators is shown in Fig. \ref{fig:laddercircconstr}, which makes use of three optimizations compared to a naive LCU implementation. First, instead of encoding the $2n$ coefficients directly in the address register preparation unitary, we only encode the $n$ unique coefficients $\vec{d}_{j}$ and we distinguish between the $Z$ and $ZX$ terms making up a ladder operator by means of a single additional ancilla qubit which is brought into an equal superposition of $\ket{0}$ and $\ket{1}$ to encode these two terms by means of a CNOT and a Z gate. Second, instead of absorbing the signs of the coefficients into the corresponding unitaries, we prepare the address register using the operator $V_{\vec{d}_j}$ which maps the all-zeroes state to $||\vec{d}_j||_1^{-1/2}\sum_{i=0}^{n-1}\mathrm{sgn}(d_{i,j})\sqrt{|d_{i,j}|}\ket{i}$ (which can be implemented using the generic method of Ref. \cite{iten2016quantum} using $\mathcal{O}(n)$ single-qubit and CNOT gates) and we uncompute the register using $V_{|\vec{d}_j|}^{\dag}$. This keeps the signs of the coefficients separated from the ladder operators, ensuring that each operator in the linear combination is an identical single-qubit unitary applied to a different target qubit. This fact is what allows our third optimization, which is to implement the operator $\sum_i\ket{i}\!\bra{i} \otimes L_i^{\dag}$ by applying a sequence of $n-1$ single-controlled SWAP gates which routes each target qubit to the top target qubit, then applying the ladder operator, and finally routing the qubit back to its original position. This is more efficient than a naive sequence of $n_{\text{addr}}$-controlled single-qubit unitaries applied to the different target qubits.

The full circuit for the approximate excited state preparation using ladder operators consists of $k$ repetitions of the circuit in Fig.~\ref{fig:laddercircconstr}, and therefore has a total gate complexity of only $\mathcal{O}(kn)$ for any state within the $k$-th excited subspace. It succeeds only if the all-zeros bit string is measured in the ancilla register for each of the $k$ rounds. In order to increase the success probability, we replace the first probabilistic LCU layer in this sequence applied to the $\ket{+}^{\otimes n}$ state with the deterministic construction of the exact first excited state, which has the same gate complexity. The approximate excited state preparation protocol is thus considerably more NISQ-friendly than the exact one, but the trade-off lies in the fact that due to its probabilistic nature it will need to be repeated a certain number of times until it is successful. As shown in Fig. \ref{fig:init_state_approx}(b), the success probabilities of the approximate preparation protocol seem to decrease roughly as $\mathcal{O}(k/n^{k-1})$, meaning that the total time complexity is roughly $\mathcal{O}(n^k)$, i.e. the same as that of the exact preparation protocol. However, a quadratic improvement compared to the exact preparation can be obtained in the approximate protocol by making use of amplitude amplification, at the cost of giving up the NISQ-friendliness. Note that the number of required repetitions in this amplitude amplification routine is known since the success probability of applying the $j$-th linear combination of ladder operators $M_j\equiv\sum_id_{i,j}L_i^{\dagger}$ to the state $\ket{b}$ obtained from all previous ladder layers is given by $(\lVert M_j\ket{b}\rVert \,/\, \lVert2\vec{d}_j\rVert_1)^2$, which can be computed classically in $\mathcal{O}(n^j)$ because in our case the state $M_j\ket{b}$ consists of the $\binom{n}{j}$ of orthonormal basis states.

\subsection{QAOA performance with problem-adjusted excited initial states}
\label{performanceoftheexcitedstateprotocol}

In Fig. \ref{fig:circ_Ising_perf}(a) we show the performance of QAOA for different initial states in terms of final ground space overlap. The results for $p=0$ show that the final ground-space overlap of the lowest level of each excited subspace $\ket{\alpha_k^0}$ increases with $k$. Two observations explain this. First, a computational basis ground state has the same overlap with every vector in $B^n_k$, so the total ground-space overlap of $X^n_k$ is guaranteed to grow in proportion to its dimension, and hence with $k$, until it reaches its maximum at $k=\lfloor n/2\rfloor$. Second, minimizing the energy within a subspace tends to correlate strongly with maximizing the ground-space overlap, so the lowest level of $X^n_k$ follows the same trend. For initial states $\ket{\alpha_k^0}$ with $k=1,2,3$ this advantage over the initial ground state (i.e. $k=0$) persists for all tested values of $p$, and we expect it to persist until they all converge to a value of 1 in the $p\rightarrow\infty$ limit. For initial states $\ket{\alpha_k^0}$ with $k=4,5,6$, however, the advantage over the initial ground state only persists for decreasingly small $p$ values as $k$ increases and eventually turns into a disadvantage for sufficiently large $p$.

The difference in performance at larger $p$ values between initial states from the first three excited subspaces and those from higher excited subspaces can be explained by means of a careful analysis of the symmetry properties of the CI Hamiltonian which, as discussed in Sec. \ref{alternativeinitialstates}, is invariant under translations, site reversal, and global spin flip. When the final ground space forms a single free orbit under these conserved symmetries, its symmetry representation, including the multiplicity of each irreducible representation, is fixed. Therefore, subject to the absence of accidental crossings between states belonging to the same irreducible representation, adiabatic connectivity can be inferred by matching the lowest-energy occurrences of these representations in the degenerate subspaces of the $X$-mixer to their multiplicities in the final ground space. As such, we can infer that the number of adiabatically connected states in the $k$-th subspace $N_k$ are given by $N_0=1$, $N_1=n$, $N_2=2n-2$, $N_3=n$, $N_4=1$, and $N_k=0$ for all $k>4$. For $k=2,3,4$ these multiplicities are smaller than the size of the corresponding subspace, and therefore this does not yet tell us for which values of $l$ the state $\ket{\alpha_k^0}$ is adiabatically connected. However, this can be determined numerically by splitting the perturbation operator restricted to the $k$-th subspace into symmetry sectors before diagonalization. We performed this symmetry analysis for a sufficiently large set of random instances for a range of problem sizes up to $n=19$. The state $\ket{\alpha_3^0}$ was adiabatically connected to the final ground space in every instance, and $\ket{\alpha_2^0}$ in most instances, with the fraction increasing with $n$. In contrast, $\ket{\alpha_4^0}$ was never observed to be connected: the single connected state for $k=4$, belonging to the zero-momentum, reflection-odd symmetry sector, typically lies high in the subspace. Therefore, aside from the decreasing number of instances for which $\ket{\alpha_2^0}$ is not connected, we may attribute the qualitative difference in performance at larger $p$ values for different values of $k$ observed in Fig. \ref{fig:circ_Ising_perf}(a) to whether or not $\ket{\alpha_k^0}$ is adiabatically connected. When this is the case, a simple monotonic QAOA parameter schedule can evolve the excited initial state into the final ground space, while preserving its advantage over the initial ground state. When the excited initial state is not connected, this is generally no longer true and the inferior performance of these initial states (at large $p$) can potentially be attributed to the inability of the classical optimization routine to find the potentially highly non-monotonic and unpredictable QAOA parameters.

The results in Fig. \ref{fig:circ_Ising_perf}(b) show that the approximation ratios paint the same qualitative picture as their corresponding final ground space overlaps, with the higher excited state advantage for $p=0$ being even more pronounced. The approximation ratios of the non-connected states drop below those of the connected states at smaller $p$ values than the corresponding ground space overlaps do. This indicates that, for comparable ground space overlap, the remaining population is distributed over higher excited states for the non-connected states than for the connected ones.
\begin{figure*}[t]
\centering
\begin{subfigure}[c]{0.475\textwidth}
\includegraphics[width=7.75cm]{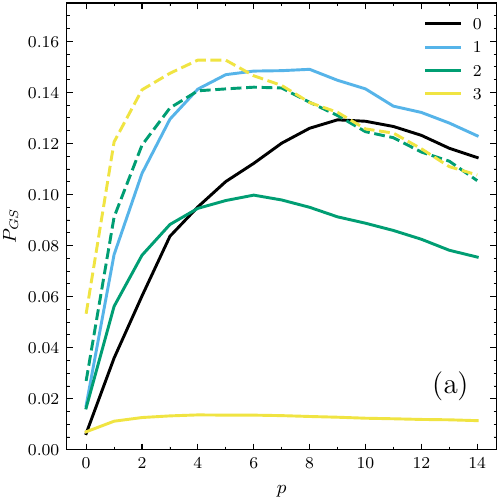}
\end{subfigure}
\hfill
\begin{subfigure}[c]{0.475\textwidth}
\includegraphics[width=7.75cm]{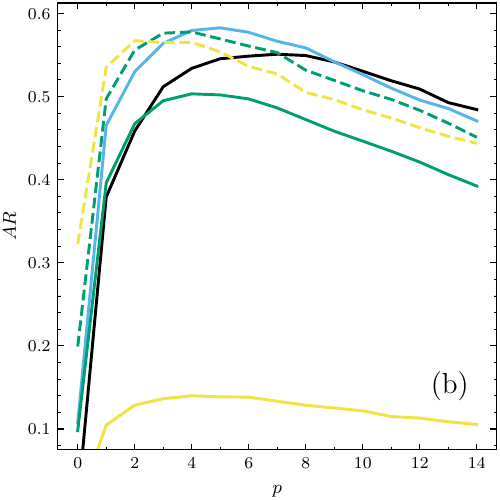}
\end{subfigure}
\caption{Mean (a) $P_{GS}$ and (b) $AR$ of fixed parameters QAOA with $p$ layers for exact initial states $\ket{\alpha_k^0}$ (solid lines) and ladder-approximated initial states $\ket{\widetilde{\alpha}_k^0}$ (dashed lines) for $k$ ranging from $0$ to $3$, averaged across $30$ instances of the CI Hamiltonian with $4n$-degenerate ground space for $n=13$, executed on a simulator with a simple noise model.}
\label{fig:noisy}
\end{figure*}

An alternative way to view the results of Fig. \ref{fig:circ_Ising_perf} is that by making use of (connected) excited initial states it becomes possible to achieve a given target ground space overlap or approximation ratio using a smaller number of QAOA layers. As such, the number of QAOA parameters which need to be classically optimized can be reduced.

In order to demonstrate that the observed advantage of the excited initial states is non-trivial, i.e. that it outweighs the increase in required quantum hardware resources, we next run experiments on a simulator with a simple noise model. The noise model consists of a depolarizing channel applied to every single- and two-qubit gate in the quantum circuit, with error probability $10^{-4}$ and $10^{-3}$, respectively. In addition, we include a symmetric qubit readout error probability of $10^{-2}$. These probabilities match the orders of magnitude of the corresponding error probabilities on currently available quantum computers. The quantum circuits are compiled down to a basis of arbitrary single-qubit gates plus the CNOT gate. Because optimizing the QAOA parameters in noisy simulations is slow and less reliable, we instead make use of the fixed parameter scheme of Ref. \cite{brandao2018fixed}. That is, we perform noiseless simulations of 30 instances and define our fixed parameters to be the mean of the resulting optimized parameters. These are then directly applied to the (separate) instances in our noisy simulations. This approach would also be suitable when considering larger problem sizes on real quantum hardware, since the fixed parameters can simply be rescaled accordingly, whereas optimizing the parameters in noiseless simulations for each corresponding problem instance becomes prohibitive.

The results of our noisy simulations are shown in Fig. \ref{fig:noisy}, clearly demonstrating how gate errors start to overwhelm the computation for sufficiently large values of $p$, as such driving the ground space probabilities and approximation ratios back down to zero. This implies that each initial state has some optimal value of $p$, at which it reaches its maximum $P_{GS}$ and/or $AR$. We see that the first excited state achieves better optimal performance than the ground state, and does so at a smaller value of $p$. The exact second and third excited states, on the other hand, achieve worse optimal performance than the ground state due to the $\mathcal{O}(n^k)$ state preparation complexity, causing the gate errors to outweigh the improved initial state advantage. However, the ladder-approximated states are able to achieve similar optimal performance as the first excited state thanks to their more efficient $\mathcal{O}(nk)$ state preparation complexity, despite the additional intermediate measurements in the LCU circuit which can cause it to prepare an incorrect state due to readout errors. As $k$ increases, the value of $p$ at which the optimum is reached decreases, which in turn reduces the required classical optimization effort. As noted earlier, the ladder-approximated state preparation requires more shots of the same circuit. However, on current NISQ devices, doing more shots is preferable over having deeper circuits.

\subsection{Using problem-adjusted initial states for general problem Hamiltonians}
\label{grovermixerinitialstates}

An alternative to the $X$-mixer Hamiltonian is the Grover mixer $H^G_M=-|\psi_0\rangle\langle\psi_0|$, where $\ket{\psi_0}$ is the equal superposition of all computational basis states (or of all feasible solutions if it is a hard-constrained problem). This Hamiltonian has a single non-degenerate ground state $\ket{\psi_0}$ with energy $-1$ and a single ($2^n-1$)-dimensional excited subspace with energy 0. The use of this mixer can recover the quadratic speedup of Grover's algorithm when used in an adiabatic evolution algorithm \cite{roland2002quantum} as well as in QAOA \cite{jiang2017near}, and for the latter it was also shown that it can outperform the $X$-mixer approach for certain problem classes \cite{bartschi2020grover}.

We generalize this Grover mixer by replacing $\ket{\psi_0}$ with our excited eigenstates $\ket{\alpha_k^0}$ of the $X$-mixer. This unlocks the power of these efficiently preparable states with favorable final ground space overlap for any problem Hamiltonian regardless of its ground space degeneracy, because the structure of the Grover mixer ensures that it will always be adiabatically connected to the final ground space. The mixer unitary is given by
\begin{equation}
    \label{eq:grovermixer}
    \begin{split}
    e^{-i\beta H^G_M} &= U_ke^{i\beta|0^n\rangle\langle0^n|}U_k^{\dag} \\
    &= U_k\left(I+(e^{i\beta}-1)|0^n\rangle\langle0^n|\right)U_k^{\dag},
    \end{split}
\end{equation}
with $\ket{0^n}\equiv\ket{0}^{\otimes n}$ and where $U_k$ is the exact state preparation unitary which prepares the lowest level of the $k$-th excited subspace of the $X$-mixer Hamiltonian, as discussed in Sec. \ref{exactstatepreparation}, and where the operator between $U_k$ and $U_k^{\dag}$ can be implemented as $X^{\otimes n}C^{(n-1)}(P(\beta))X^{\otimes n}$, with arbitrary assignment of control qubits and target qubit for the multi-controlled phase operator.

\begin{figure*}[t]
\centering
\begin{subfigure}[c]{0.475\textwidth}
\includegraphics[width=7.75cm]{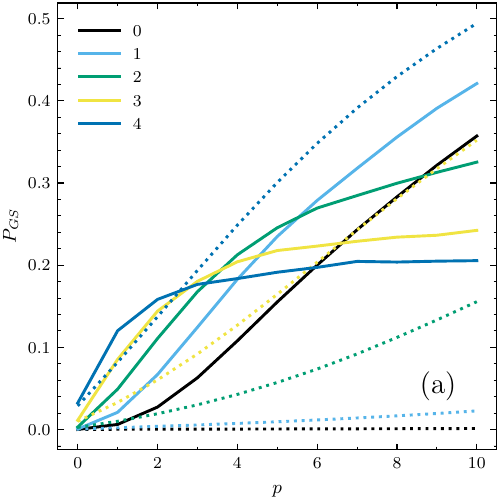}
\end{subfigure}
\hfill
\begin{subfigure}[c]{0.475\textwidth}
\includegraphics[width=7.75cm]{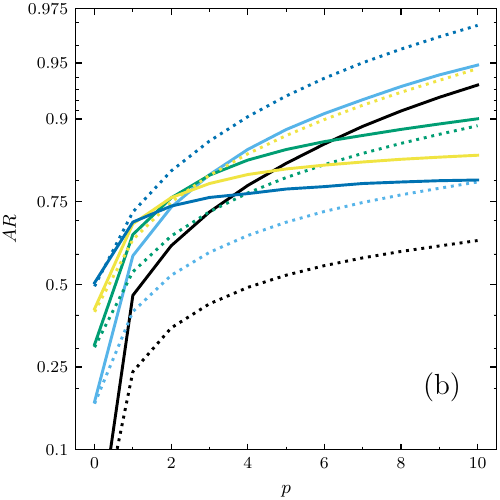}
\end{subfigure}
\caption{(a) Mean $P_{GS}$ of QAOA with $p$ layers with $X$-mixer Hamiltonian (solid lines) and Grover-mixer Hamiltonian (dotted lines) for initial states $\ket{\alpha_k^0}$ for $k$ ranging from $0$ to $4$, averaged across 50 instances of the $n$-qubit SK Hamiltonian for $n=15$. (b) Corresponding mean $AR$ for the same experiments (with $Y$ axis in logit scale for improved readability).}
\label{fig:grover}
\end{figure*}

In Fig. \ref{fig:grover} we compare the performance of the $X$-mixer and the Grover mixer when applied to the SK Hamiltonian. Recall that this problem class has a two-fold degenerate ground space generated only by the global spin-flip symmetry. As a result, only the ground state and the lowest level of the first excited subspace are connected to the final ground space, whereas higher excited states are not. This is clearly reflected in the results obtained with the $X$-mixer, for which the first excited state consistently outperforms the ground state, while the second, third, and fourth excited states only outperform the ground state for small values of $p$ and actually underperform compared to the ground state for larger values of $p$. When using the Grover mixer, the results obtained using the ground and first excited states are worse compared to their counterparts using the $X$-mixer. However, now the second, third, and fourth excited states consistently outperform the ground and first excited states, and, for the fourth excited state and sufficiently large $p$, even any result obtained using the $X$-mixer. This clearly illustrates the main takeaway of this Subsection, which is that by using the Grover mixer we can bypass the restrictions imposed by the degeneracy of the target Hamiltonian. As such, our excited state QAOA can be applied to any problem class, including problems with varying degeneracy, no degeneracy, or unknown degeneracy, and as such obtain superior performance compared to standard QAOA. It should be noted that the Grover mixer layer has a considerably higher $\mathcal{O}(n^k)$ gate complexity than the constant-depth $X$-mixer layer and therefore, given a fixed total quantum gate budget, one could use many more QAOA layers for the $X$-mixer and as such outperform the corresponding Grover mixer circuit. However, this would come at the cost of having to do more classical optimization work to optimize the QAOA parameters.

\section{Conclusion}
\label{conclusion}

In summary, we have discussed how the use of problem-adjusted superpositions of degenerate excited eigenstates of the $X$-mixer Hamiltonian can improve the performance of QAOA both in the absence and presence of quantum gate and readout noise. We have explained how the correct superposition can be found by means of degenerate perturbation theory, and we have demonstrated how it can be efficiently prepared with $\mathcal{O}(n^k)$ quantum gate complexity. In addition, we have presented an approximate protocol for preparing these states using a linear combination of ladder operators. This protocol can be used in a repeat-until-success scheme to reduce the quantum gate complexity to $\mathcal{O}(kn)$, at the cost of $\mathcal{O}(n^{k-1})$ repetitions, or it can be used in an amplitude amplification scheme to obtain an end-to-end quadratic improvement over the exact state preparation protocol. Our numerical simulations showed remarkably good overlap between the approximate and the exact excited states.

The advantage of the $k$-th excited state over the ground state of the $X$-mixer Hamiltonian is to be expected because it minimizes the expectation value of the problem Hamiltonian $H_P$ within an ${n \choose k}$-dimensional subspace, as such achieving increasingly good overlap with the ground space of $H_P$ as $k$ grows (up to $\lfloor n/2 \rfloor$). When used as an initial state in QAOA, the picture becomes more nuanced. In particular, the advantage of the $k$-th excited state only persists for increasing values of $p$ if it has a reasonable time evolution path into the ground space of $H_P$, potentially because this property is what makes it feasible for the classical optimization routine to find sufficiently good variational parameters in reasonable time. Since the highest possible excited initial state for which this is the case can depend on the specific instance of $H_P$ and may not be known a priori, this can put a limitation on the practical use of the excited initial state approach. In order to address this, we introduced a variant of QAOA in which an excited initial state of the $X$-mixer is inserted into a Grover-like mixer Hamiltonian. This ensures that the chosen state is guaranteed to be the ground state of the new mixer Hamiltonian and therefore that it adiabatically evolves into the ground space of $H_P$, regardless of the degeneracy properties of $H_P$. This generically unlocks the power of these improved initial states for any problem Hamiltonian instance and indeed our numerical simulations confirmed that this approach significantly outperforms standard QAOA.

It is worth noting that, even though we applied these improved initial states to diagonal problem Hamiltonians, our formalism and techniques can just as well be applied to obtain improved approximations to the ground state of non-diagonal Hamiltonians, such as for certain spin problems or quantum chemistry problems. In addition, for such problem Hamiltonians it may also be of interest to (approximately) construct excited states of $H_P$, for which our scheme also provides a convenient pathway.

We conclude with two final remarks about our proposed excited initial states. First, they do not avoid the quantum Monte Carlo sign problem because any excited $X$-mixer eigenstate is orthogonal to $\ket{+}^{\otimes n}$ and therefore necessarily contains both positive and negative computational basis amplitudes, preventing their direct interpretation as non-negative sampling weights. For the Grover-style mixer approach, this sign structure is additionally inherited by the Hamiltonian itself, making it non-stoquastic in the computational basis. Second, in contrast to the ground state of the $X$-mixer they are entangled states. As such, the analysis presented in \cite{farhi2014quantum} based on counting subgraph types for fixed (small) $p$ values cannot be used, and the approximation ratio results which were found in that paper therefore do not apply to our proposed QAOA variant, which in general outperforms standard QAOA.

\section*{Disclaimer}
This paper was prepared for information purposes and is not a product of HSBC Bank Plc. or its affiliates. Neither HSBC Bank Plc. nor any of its affiliates make any explicit or implied representation or warranty and none of them accept any liability in connection with this paper, including, but not limited to, the completeness, accuracy, reliability of information contained herein and the potential legal, compliance, tax or accounting effects thereof.

\bibliographystyle{unsrt}
\bibliography{refs}

\end{document}